\documentclass[aps,prd,12point,twocolumn,nofootinbib,showpacs,superscriptaddress]{revtex4-1}

\usepackage{amssymb}
\usepackage{graphicx}
\usepackage{amsmath, amsthm}
\usepackage{epstopdf}
\usepackage{hyperref}

\usepackage{amsmath,amsfonts,amssymb,mathrsfs}
\usepackage{graphicx}
\usepackage{color}
\def\be{\begin{equation}}
\def\ee{\end{equation}}
\begin{document}
\title{Beyond lensing by the cosmological constant}

\author{Valerio Faraoni}
\email{vfaraoni@ubishops.ca}
\affiliation{Physics Department and STAR Research Cluster, 
Bishop's University, 2600 College Street, Sherbrooke, 
Qu\'ebec, Canada J1M 1Z7}

\author{Marianne Lapierre-L\'eonard}
\email{mlapierre12@ubishops.ca}
\affiliation{Physics Department, 
Bishop's University, 2600 College Street, Sherbrooke, 
Qu\'ebec, Canada J1M 1Z7}

\begin{abstract}
The long-standing problem of whether the cosmological constant 
affects directly the deflection of light caused by a 
gravitational lens is reconsidered. We use a new approach based 
on the Hawking quasilocal mass of a sphere grazed by light rays 
and on its splitting into local and cosmological parts. 
Previous literature restricted to the cosmological constant is 
extended to any form of dark energy accelerating the
 universe in which the gravitational lens is embedded.
\end{abstract}

\maketitle

The deflection of light by a localized mass is one of the three 
classical tests of General Relativity \cite{Wald, Willbook} 
and the verification of Einstein's prediction for the 
deflection angle of light rays grazing the Sun by Eddington 
\cite{Eddington} during the 1919 
solar eclipse turned Einstein into a celebrity.  Following 
the discovery of the 
first gravitational lens system 0957+561 in 1979 
\cite{Walshetal},  
gravitational lensing rapidly developed 
to become a major tool of cosmology and astrophysics, 
providing information about the abundance and distribution of 
dark matter at various scales, about microlensing by stars and 
(exo-)planets, and revealing gravitational 
lens systems acting as giant telescopes  
\cite{FalcoSchneider, Perlick}. In addition to 
lensing by localized mass 
distributions, in the 1980s researchers started to enquire 
about the direct contribution to lensing by the cosmological 
constant $\Lambda$. Today, it is only natural to 
investigate further this subject, since we now know that 
the present cosmic 
expansion accelerates due to a true or an effective 
cosmological constant \cite{tutti, Lake, Sereno, Heavens, 
Aghili}. The mystery of what propels the acceleration of the cosmic 
expansion and the fact that the cosmological constant
 energy density (the density of quantum vacuum) 
is 120 orders of magnitudes smaller than what predicted 
({\em i.e.}, the cosmological constant problem) 
are two major unsolved 
puzzles of modern theoretical physics \cite{AmendolaTsujikawa, 
Weinberg}.
During the last decade  
there has been an active debate about the 
direct effect of the 
cosmological constant in lensing, with various approaches 
often leading to opposite results (although sometimes these 
are due to different definitions of the physical 
quantities involved, see {\em e.g.} 
Ref.~\cite{Lake}). The debate is not  
settled and here we contribute a 
different 
approach which is more general, simple yet powerful, and 
covariant. 
Almost all the existing works on the subject are restricted 
to studying lensing in a Schwarzschild-de Sitter/Kottler 
spacetime with a central mass $m$ and the cosmological 
constant $\Lambda$ \cite{Kottler}, to a vacuole in the 
Kottler geometry, or (rarely) to the 
McVittie geometry \cite{McVittie}. The Kottler spacetime 
is (locally) static and, therefore, 
very special and the McVittie solution of the Einstein 
equations is also special and may be misleading when 
drawing general conclusions. It is preferable to study 
instead lensing by 
a localized mass distribution in {\em any} 
realistic (accelerating or, in principle, even 
decelerating) Friedmann-Lema\^itre-Robertson-Walker (FLRW) 
universe perturbed by local mass distributions acting as 
lenses.  The statement that the cosmological constant 
does or does not contribute directly to light deflection 
ultimately relies on splitting the lensing 
mass-energy enclosed in a sphere 
grazed by the light rays into a local (Newtonian) part 
which is attributed to the lens and 
a cosmological contribution due to the cosmological constant 
or to the cosmic fluid. However, this decomposition issue is 
left implicit in the existing investigations of this subject 
because the formalisms used are not suitable for tackling 
this problem. 
Here we address explicitly the notion of 
``mass'' deflecting light rays, keeping in mind that the mass 
of  
a gravitating system is a nontrivial concept in 
non-asymptotically flat spacetimes describing non-isolated lenses,  
which are necessarily embedded in the universe 
(taking into account the contribution of $\Lambda$ means,
 by definition, that the lens is not asymptotically flat). 
This fact has 
led to various quasilocal energy definitions (see 
Ref.~\cite{Szabados} 
for a recent review). The total mass-energy of  
a gravitating system includes contributions from rest 
mass, kinetic and potential energies, and the energy of 
the gravitational field, which itself gravitates. The 
essence of the  Equivalence 
Principle constituting the foundation of General 
Relativity and of all metric theories of gravity 
\cite{Wald, Willbook} is that the gravitational field 
can be eliminated locally. Therefore, to this order, one 
cannot introduce a local energy density for the 
gravitational field. The next best thing is to 
define the 
total energy of matter and fields enclosed by a compact, 
spacelike  2-surface, {\em i.e.}, quasilocally, and there 
are several possible definitions of quasilocal mass 
\cite{Szabados}. Here we adopt the 
Hawking-Hayward 
quasilocal energy construct \cite{Hawking, Hayward} which, 
in spherical 
symmetry (to which, although not necessary,  we restrict 
for simplicity following the previous literature on the subject), 
reduces to the better known 
Misner-Sharp-Hernandez mass \cite{MSH} widely used in 
relativistic fluid mechanics and in black hole 
thermodynamics. The original definition of Hawking 
\cite{Hawking} (to which Hayward later added a term \cite{Hayward}) 
is based on a two-dimensional spacelike, compact, embedded 
hypersurface ${\cal S}$ and on integrating over ${\cal S}$ the 
squares of the expansion and shear of the  outgoing and 
ingoing null geodesic congruences from ${\cal S}$. Therefore, the 
basic idea is that the Hawking quasilocal construct weights the  
total mass in ${\cal S}$ by using its effect on light 
rays, which fits well the lensing problem that we consider 
here. This mass concept 
has been applied recently to the study of two problems in 
cosmology, Newtonian $N$-body simulations of large scale structure 
formation \cite{Nbody} and the turnaround radius of  
a large structure in an accelerating universe \cite{turnaround}. 
The Hawking-Hayward quasilocal mass splits uniquely 
into three contributions: a local one, a cosmological one, 
and a (much smaller) one describing the interaction between the 
previous 
two. This feature is well suited for the discussion of   
lensing by the cosmological constant or by the dark energy 
propelling the accelerated expansion of the cosmos.  

We begin by restricting ourselves to General 
Relativity and by considering a spherically symmetric  
perturbation of a spatially flat FLRW universe, which is 
localized in a 
region of size much 
smaller than the Hubble radius $H^{-1}$ and deflects 
light rays propagating nearby. The usual thin lens and 
small deflection angle approximations \cite{FalcoSchneider, 
Perlick} are adopted here. We use units in which the speed 
of light $c$ and Newton's constant $G$ are unity. 
The line element in the conformal Newtonian 
gauge is 
\begin{eqnarray} 
d\tilde{s}^2&=& a^2(\eta) \left[ 
-\left( 1+2\phi \right) d\eta^2 +\left( 1-2\phi \right) 
\left( dr^2 +r^2 d\Omega_{(2)}^2 \right) \right] \nonumber\\
&&\nonumber\\
& \equiv & a^2 ds^2\,,  \label{CNG}
\end{eqnarray}
where the scale factor $a(\eta)$ is a function 
of the conformal time $\eta$, $\phi(r)=-m/r $ describes the 
local spherical perturbation with Newtonian mass $m$, and 
$d\Omega_{(2)}^2=d\theta^2 +\sin^2 \theta \, d\varphi^2$ is 
the metric on the unit 2-sphere. Working to first order and 
following standard literature, we  
do not include vector and tensor 
perturbations in the line element~(\ref{CNG}).  Vector and tensor 
perturbations would not 
be negligible to second order due to mode-mode coupling, but they 
can be safely neglected to first order for gravitational lenses 
which do not have relativistic peculiar 
velocities \cite{Mukhanov, perts}.\footnote{Lensing by 
astrophysical gravitational waves (tensor modes) is an ephemeral 
phenomenon which was studied, {\em e.g.}, in~\cite{gwlensing} and  
has never been observed in nature.} Moreover, we restrict to 
spherical scalar perturbations: 
this assumption is not strictly necessary, but it greatly 
simplifies the discussion and allows one to speak of ``the 
deflection angle'' in the same way that ``the deflection angle by 
the Sun'' was calculated by Einstein and appears in relativity 
textbooks. Moreover, spherically symmetric gravitational lenses are 
used almost universally in the 
literature on lensing contributions by the cosmological constant.
 
The Hubble function is $H(t)=\dot{a}/a$, where an overdot 
denotes differentiation with respect to the comoving time 
$t$ of the FLRW background, which is related to the 
conformal time $\eta$ by $dt=ad\eta$. 
We denote with a tilde geometric quantities in the 
perturbed FLRW space~(\ref{CNG})  
obtained by means of a conformal transformation of the 
linearized Schwarzschild metric (usually written as  
$g_{ab} \rightarrow \tilde{g}_{ab}=\Omega^2 g_{ab}$)  
\be \label{linearizedSchwarzschild} 
ds^2= -\left( 1+2\phi 
\right) d\eta^2 +\left( 1-2\phi \right) \left( dr^2 +r^2 
d\Omega_{(2)}^2 \right) 
\ee 
with conformal factor 
$\Omega=a(\eta)$. A possible time-dependence of the perturbation 
potential $\phi$ is neglected since the latter describes  
a Newtonian lens affected very little by the cosmological 
expansion over the time during which lensing takes place (lensing 
by the localized mass $m$ only occurs when 
the light rays are near it).  
The reason to write explicitly 
the line element~(\ref{CNG}) as the result of a conformal   
rescaling is that null 
geodesics are 
conformally invariant and angles are left invariant by 
conformal transformations. The deflection angle of light 
rays in Schwarzschild space (in the small angle 
approximation) is 
\be \label{deflection} 
\Delta \varphi= \frac{4m}{r}  
\ee 
to first order in the Newtonian 
potential $\phi=-m/r$, where $r$ 
is the impact parameter. The deflection angle will be the 
same in the conformally rescaled spacetime~(\ref{CNG}), 
$\Delta 
\tilde{\varphi}=\Delta \varphi$.

The second ingredient of our new approach consists of 
noting that a light ray  
deflected at (areal) radius $R$ ``sees'' the entire 
physical mass contained in a sphere of radius $R$. (Here we 
employ the areal, instead of coordinate, radius 
because the former is a geometric quantity defined independently 
of the coordinate system.)  This 
mass is given by the Hawking-Hayward/Misner-Sharp-Hernandez 
construct $M_\text{MSH}$ \cite{MSH}. The latter is defined, in 
General 
Relativity and in a spherically symmetric spacetime with 
areal radius $R$, by \cite{MSH, Haywardspherical} 
\be \label{MSH} 
1-\frac{2M_\text{MSH}}{R}=\nabla^c R \nabla_c R \,. 
\ee 
This scalar equation and the fact that $R$ is a 
geometrically defined quantity in any spherically 
symmetric spacetime make it clear that $M_\text{MSH}$ 
is defined in a geometric and, therefore,  gauge-invariant 
way. 
In the Schwarzschild spacetime,  $M_\text{MSH}$ coincides with 
the Schwarzschild mass $m$ but, in the conformally rescaled 
FLRW space~(\ref{CNG}), it receives contributions by the 
cosmological fluid (which, possibly, reduces to just the   
cosmological constant) and by the gravitational field. 
The 
transformation property of the Hawking-Hayward mass under 
conformal transformations $g_{ab} \rightarrow 
\tilde{g}_{ab}= \Omega^2 g_{ab}$, derived  in 
\cite{hhconfo, Enzo2013}  
and applied to perturbed FLRW spaces in \cite{Nbody, turnaround},  
is 
\be\tilde{M}_\text{MSH}= \Omega M_\text{MSH}-\frac{R^3}{2\Omega} \,
 \nabla^c \Omega \nabla_c\Omega -R^2 \nabla^c \Omega \nabla_c R \,,
\ee
which in our case yields
\begin{eqnarray}  
\tilde{M}_\text{MSH} &=& 
m\, a(t) +\frac{H^2 \tilde{R}^3}{2} \left( 1-2\phi \right) \\
&&\nonumber\\
& \simeq & 
m \, a(t) +\frac{H^2 \tilde{R}^3}{2}  \,, 
\label{MSHtransform}
\end{eqnarray} 
expressed in terms of the comoving FLRW time $t$ and of the 
areal radius of the perturbed FLRW space~(\ref{CNG}) 
\be\label{tildeR} 
\tilde{R}=a(t) r \sqrt{1-2\phi} \simeq ar 
\left( 1-\phi \right)  \,.
\ee 
Our quantification of the direct 
contribution of the cosmological fluid to lensing hinges on the 
decomposition~(\ref{MSHtransform}) of the gravitating mass 
\cite{Nbody, Enzo2013} 
into a ``local'' part $m \,a(t)$, a ``cosmological'' part $ 
H^2  \tilde{R}^3/2$, and an ``interaction'' part $ - H^2
\tilde{R}^3 \phi$. It may seem surprising that the 
contribution of the Newtonian 
mass $m$ to $\tilde{M}_\text{MSH}$ scales with $a(t)$, but 
this is 
not so strange if the local mass is regarded as a length 
scale (in units in which $G=c=1$ it is one half of the 
Schwarzschild radius of the mass $m$, which is commonly 
regarded as a length scale in relativistic astrophysics). 
The second contribution to 
$\tilde{M}_\text{MSH}$ can be written as $ H^2
\tilde{R}^3/2 = 4\pi \tilde{R}^3 \rho /3$ by virtue of the 
Hamiltonian constraint
\be
H^2 =\frac{ 8\pi}{3} \, \rho \, \dot{=} \, \frac{\Lambda}{3} \,,
\ee
where $\rho$ is the energy density of the cosmic fluid and the 
last equality is characterized by the symbol $\dot{=}$  
denoting the fact that it holds  
only in the special case in which the cosmic fluid is composed 
solely of a cosmological constant $\Lambda$ with density $\Lambda/( 
8\pi )$. 
The cosmological contribution to $\tilde{M}_\text{MSH}$ is 
simply the mass of cosmic fluid enclosed by a sphere of 
radius $\tilde{R}$ \cite{Hayward}. The deflection angle will 
receive 
corresponding contributions of 
magnitudes $\tilde{M}_\text{MSH}/\tilde{R}$ and 
$H^2\tilde{R}^2$, which can already be compared. They are 
both small in the gravitational lens systems usually 
considered (stars, galaxies, or galaxy clusters) but the 
second term (which is quadratic in the ratio lens size$/$Hubble 
radius $H^{-1}$) is much smaller than the 
first one (which is linear in the ratio Schwarzschild 
radius of the 
lens$/$size of the system). The interaction term 
$ - H^2\tilde{R}^3 \phi$ is naturally smaller than both and 
is included here only for comparison with previous literature. 
To illustrate the magnitude of these contributions consider, 
for example, a galaxy with mass $m\sim 10^{11} M_{\odot}$ and size
$R\sim 25 $~kpc at redshift $z \simeq 1$. Then, using $a_0/a=z+1$ 
and adopting the usual convention $a_0=1$ with the value 
$H_0=70 \, \mbox{km}/(\mbox{s} \cdot \mbox{Mpc}) $ (and restoring 
Newton's constant $G$ and the speed of light $c$), one has 
\begin{eqnarray}
\frac{H^2 \tilde{R}^3}{2c^2} \simeq 1.3 \cdot 10^{10} \, \mbox{m} 
\,,\\
&&\nonumber\\
\frac{Gma}{c^2} \simeq 7.4 \cdot 10^{13} \, \mbox{m} \,,
\end{eqnarray}
and their ratio is 
\be
\frac{ H^2 \tilde{R}^3/2c^2 }{ Gma/c^2} \simeq 2 \cdot 10^{-4} 
\,,
\ee
hence the mass $\tilde{M}_\text{MSH}$ in eq.~(\ref{MSHtransform}) 
is completely dominated by the contribution due to the local 
gravitational lens mass and the contribution of the cosmological 
fluid is negligible in comparison.

We now replace ratios $m/r$ in the deflection angle $\Delta 
\tilde{\varphi}=\Delta \varphi$ (eq.~(\ref{deflection})) 
with $ma/ar$ and we substitute the expressions of $ma$ and 
$ar$ obtained from eqs.~(\ref{MSHtransform}) and 
(\ref{tildeR}) obtaining
\be
\Delta\tilde{\varphi} = \frac{4ma}{ar}  
=  \frac{4\left[ 
\tilde{M}_\text{MSH} 
-\frac{H^2 \tilde{R}{^3}}{2} \left( 1-2\phi \right) 
\right]}{\tilde{R} \left( 1+ \phi \right)} 
 \,.\label{questa}
\ee
Since both $ | \phi| \ll 1$ and $H^2 \tilde{R}^2 \ll 1$, 
a first order expansion in these quantities yields
\be\label{final}
\Delta \tilde{\varphi} = \frac{ 4 \tilde{M}_\text{MSH}}{ 
\tilde{R}} -2 H^2 \tilde{R}^2 \,.
\ee
This is our main result. If the  
expression~(\ref{MSHtransform}) of $\tilde{M}_\text{MSH}$ is 
reintroduced into the deflection angle~(\ref{questa}), the  
contributions to $ \Delta \tilde{\varphi}$ due to the cosmological 
background cancel out exactly to leave $ \Delta 
\tilde{\varphi}=4m/r$. By taking the limit $m\rightarrow 0$ 
in $\Delta \tilde{\varphi}$, the deflection angle vanishes. 
The first non-vanishing contribution due to the cosmology is  
$+H^2 
\tilde{R}^2 \phi$ which, using the numerical example above, 
contributes only a fraction $\sim 6.5 \cdot 10^{-18}$ of the 
deflection angle caused by the local mass $m$.

In comparison with other 
calculations in the literature, our derivation is 
straightforward thanks to the conformal invariance of 
the deflection angle and to the  previous derivation of the 
result~(\ref{MSHtransform}). If the total gravitating 
mass $\tilde{M}_\text{MSH}$ contained in a sphere of 
radius $\tilde{R} \simeq a(t)r$ is considered, 
eq.~(\ref{final}) seems 
to imply that the cosmological constant (or, more  
generally, dark energy) contributes to 
light deflection. In the special case of the Kottler metric 
for which $H^2 = \Lambda/3$ (but keep in mind that the 
result~(\ref{final}) is more general since it applies to any 
spatially flat FLRW background, not only to de Sitter), 
eq.~(\ref{final}) 
would lead to a 
contribution $-2\Lambda \tilde{R}^2 /3$ deflecting light in 
the opposite direction of the deflection induced by the Newtonian mass 
$m$. However, when 
one splits the mass $\tilde{M}_\text{MSH}$ of the first 
term $4\tilde{M}_\text{MSH}/\tilde{R}$  into local, 
cosmological, and interacting contributions according 
to eq.~(\ref{MSHtransform}), the cosmological contribution 
cancels exactly the term $-2H^2 \tilde{R}^2$, as described 
by eq.~(\ref{questa}), hence it appears that the cosmological constant
$\Lambda$, or the cosmic fluid, or their combination, do 
not contribute directly to lensing. Therefore, the statement 
that the cosmology (or, less generally, the cosmological constant 
$\Lambda$) does not contribute directly to 
light deflection depends on 
whether one wants to identify the lens mass with the Newtonian 
mass $m$ or with the 
total physical mass $\tilde{M}_\text{MSH}$ enclosed by a sphere 
centered on the spherical lens and radius equal to the impact parameter. 
Sharp statements about the contribution of the cosmology to lensing 
should not be made without specifying this choice, but previous 
literature has ignored this aspect. 
We believe that one should include in the description of lensing 
the physical quasilocal mass, instead of the Newtonian mass
of the lens, which is appropriate only when the latter is 
isolated and this is not the case when one wants to study 
the contribution of the cosmological background to lensing. 
In this light, the result~(\ref{final}) should 
clarify much of the debate 
in the literature \cite{tutti, Lake, Sereno, Heavens,Aghili}. 
The existing debate seems to be due to the fact that different 
authors actually pose different questions and do not investigate 
a problem formulated uniquely in an unambiguous way.

Finally, we comment on the use of the McVittie metric in 
the literature on lensing by $\Lambda$ \cite{Piattella, 
Aghili}. The McVittie metric 
\cite{McVittie} is a spherically symmetric solution of 
the Einstein equations interpreted as describing a 
spherical inhomogeneity  embedded 
in a FLRW universe. It 
interpolates between the Schwarschild and the FLRW 
geometries and it contains the Kottler solution as a 
special case \cite{Arakida}, but it 
is time-dependent and more general. However, it cannot 
play a role analogous 
to that which the 
Schwarzschild space plays among asymptotically flat 
solutions of the vacuum Einstein equations. While the Schwarzschild 
geometry is the unique 
spherically symmetric, vacuum, asymptotically flat solution 
of the Einstein equations \cite{Wald}, there is no unique 
spherically 
symmetric, asymptotically FLRW solution of the Einstein 
equations. Using the areal radius $R$, the McVittie metric 
can be written as \cite{Roshina}
\begin{eqnarray}
ds^2 &=& - \left( 1-\frac{2m}{R} -H^2(t)R^2 \right) dt^2 
-\frac{2 H(t) R}{ \sqrt{ 1-2m/R}} \, dt dR \nonumber\\
&& \nonumber\\
&\, & +\frac{dR^2}{1-2m/R} + R^2 d\Omega_{(2)}^2  
\,,\label{McVittie}
\end{eqnarray}
where $m$ is a constant. It reduces to the Kottler metric 
when $H=$~const. The Misner-Sharp-Hernandez mass 
contained in  a sphere of areal radius $R$ in this metric 
is 
calculated from eq.~(\ref{MSH}) as
\be\label{MSHmcvittie}
M_\text{MSH}^\text{(McV)} = m +\frac{H^2 R^3}{2} 
= m +\frac{4\pi  R^3}{3} \, \rho  \,.
\ee
This mass splits cleanly into local and cosmological parts, 
although the McVittie metric is not conformal to 
Schwarzschild and eq.~(\ref{MSHmcvittie}) is not computed 
using eq.~(\ref{MSHtransform}) but is derived directly from 
the definition of Misner-Sharp-Hernandez mass (\ref{MSH}). 
There is no local-cosmological interaction term in this 
exact expression and one could be led to believe that such 
an interaction term is absent in more realistic situations. 
More important, the local mass contribution to the McVittie 
geometry is constant ($m$ instead of $m 
a(t)$),\footnote{This happens because, when the McVittie 
line element is written in isotropic coordinates in analogy 
with the perturbed FLRW of eq.~(\ref{CNG}), the quantity 
$m/ar$ appears instead of $m/r$ \cite{McVittie}.} and there 
is a spacelike singularity at a finite radius which, 
depending on the behaviour of the scale factor, is covered 
by a time-dependent apparent horizon. While, realistically, 
lensing is much more common in the weak-field than in the 
strong-field regime and the interpretation of the McVittie 
apparent horizon becomes less important, one is still using 
a very special solution of the Einstein equations to draw 
general conclusions, and the issue of constant $m$ versus 
$m \, a(t)$ is at the core of the direct contribution of the 
cosmological constant to the deflection angle.  The use of 
the perturbed FLRW space is more appropriate than that of 
special solutions of the Einstein equations to assess this 
contribution to lensing.

{\em Acknowledgments:} We are grateful to Bishop's 
University and to the Natural Sciences and Engineering 
Research Council of Canada for financial support.

\end{document}